\documentclass[english,aps,prx,reprint,superscriptaddress,longbibliography]{revtex4-2}

\usepackage[T1]{fontenc}
\usepackage[utf8]{inputenc}
\usepackage{amssymb}
\usepackage{graphicx}
\usepackage{subscript}
\usepackage{amsmath}
\usepackage[colorlinks=true,citecolor=blue,urlcolor=blue,linkcolor=blue]{hyperref}
\usepackage{nameref}
\usepackage[export]{adjustbox}
\usepackage{ragged2e}
\usepackage{comment}

\usepackage{subcaption}
\usepackage{caption}
\makeatletter
\usepackage{pslatex}
\newcommand{\figpart}[2]{Figure~\hyperref[#1]{\ref*{#1}#2}}
\usepackage{hyperref}
\usepackage{xcolor}
\hypersetup{
    colorlinks,
    linkcolor={blue!70!black},
    citecolor={blue!70!black},
    urlcolor={blue!70!black}
}

\usepackage{babel}

\newcommand{\X}{\tilde{X}^{2}\Sigma^{+}}
\newcommand{\A}{\tilde{A}^{2}\Pi_{1/2}}
\newcommand{\B}{\tilde{B}^{2}\Sigma^{+}}
\newcommand{\wn}{~\rm{cm}^{-1}}

\newcommand{\appropto}{\mathrel{\vcenter{
  \offinterlineskip\halign{\hfil$##$\cr
    \propto\cr\noalign{\kern2pt}\sim\cr\noalign{\kern-2pt}}}}}

\newcommand{\OverallDetuning}{165} 
 
\newcommand{\BeamSize}{1}

\newcommand{\FactorIncrease}{12}

\newcommand{\NewODT}{2.2} 
\newcommand{\NewODTUncert}{3} 
\newcommand{\NewMOT}{3.8} 
\newcommand{\NewMOTUncert}{5} 
\newcommand{\TCPower}{1}
\newcommand{\TCBeamsratio}{25}
\newcommand{\TCIntensity}{2.2}
\newcommand{\ODTPeakDensity}{2}
\newcommand{\ODTPeakDensityUnc}{1}
\newcommand{\Beta}{4}

\newcommand{\TCIoverIsat}{740}
\newcommand{\SrOHIoverIsat}{3}
\newcommand{\ODTvol}{2.1}

\newcommand{\ODTHold}{150} 
\newcommand{\ODTPrediction}{5}

\newcommand{\ODTPredictionPlus}{2}
\newcommand{\ODTPredictionHold}{40}
\newcommand{\SingleBody}{1.9}
\newcommand{\SingleBodyUnc}{1} 
\newcommand{\ODTwaist}{39.8} 
\newcommand{\ODTpower}{18.4} 
\newcommand{\ODTtemp}{54} 
\newcommand{\ODTtempunc}{1} 
\newcommand{\ODTtrapdepth}{0.5}
\newcommand{\radialwidth}{6.5} 
\newcommand{\axialwidth}{1.1}

\begin{document}

\makeatother

\title{Order of Magnitude Improved Optical Trapping of Molecules Through Transverse Cooling}

\author{Abdullah Nasir}
\affiliation{Harvard-MIT Center for Ultracold Atoms, Cambridge, Massachusetts 02138, USA}
\affiliation{Department of Physics, Harvard University, Cambridge, Massachusetts 02138, USA}

\author{Annika Lunstad}
\affiliation{Harvard-MIT Center for Ultracold Atoms, Cambridge, Massachusetts 02138, USA}
\affiliation{Department of Physics, Harvard University, Cambridge, Massachusetts 02138, USA}

\author{Mingda Li}
\affiliation{Harvard-MIT Center for Ultracold Atoms, Cambridge, Massachusetts 02138, USA}
\affiliation{Department of Physics, Harvard University, Cambridge, Massachusetts 02138, USA}

\author{Saif Salim}
\affiliation{Harvard-MIT Center for Ultracold Atoms, Cambridge, Massachusetts 02138, USA}
\affiliation{Department of Physics, Harvard University, Cambridge, Massachusetts 02138, USA}

\author{Shuqi Liu}
\affiliation{Harvard-MIT Center for Ultracold Atoms, Cambridge, Massachusetts 02138, USA}
\affiliation{Department of Physics, Harvard University, Cambridge, Massachusetts 02138, USA}

\author{Christian Hallas}
\affiliation{Harvard-MIT Center for Ultracold Atoms, Cambridge, Massachusetts 02138, USA}
\affiliation{Department of Physics, Harvard University, Cambridge, Massachusetts 02138, USA}

\author{Zack Lasner}
\altaffiliation{Current address: IonQ, Inc., College Park, Maryland 20740, USA}
\affiliation{Harvard-MIT Center for Ultracold Atoms, Cambridge, Massachusetts 02138, USA}
\affiliation{Department of Physics, Harvard University, Cambridge, Massachusetts 02138, USA}

\author{John Doyle}
\affiliation{Harvard-MIT Center for Ultracold Atoms, Cambridge, Massachusetts 02138, USA}
\affiliation{Department of Physics, Harvard University, Cambridge, Massachusetts 02138, USA}

\date{\today}

\begin{abstract}

We demonstrate a two-dimensional Sisyphus laser cooling method that increases the number of strontium monohydroxide (SrOH) molecules loaded into a magneto-optical trap by a factor of 12. Subsequent loading into an optical dipole trap (ODT) achieves $\NewODT (\NewODTUncert)\times10^4$ ultracold SrOH molecules with a peak density of $\sim\ODTPeakDensity(\ODTPeakDensityUnc)\times10^{10}~\mathrm{cm^{-3}}$. The lifetime of molecules in the ODT is limited by two-body collisions characterized by a measured collision rate constant $\beta \sim \Beta\times10^{-10}~\mathrm{cm^3/s}$. The cooling method developed here is generally applicable to all known cases of direct molecular laser cooling, including symmetric and asymmetric top molecules. Increases in trapped molecule number will directly improve the search for ultralight dark matter, position polyatomic molecules as a platform for probing CP-violating new particles with masses $\gg$10 TeV, and facilitate a broad range of further research in quantum science. 

\end{abstract}

\maketitle

\section{Introduction} The complex rovibrational structure of molecules makes them a valuable platform for a wide range of physics, including quantum computing~\cite{demille2002quantum, yu2019scalable, picard2025entanglement}, quantum simulation~\cite{Wall2013}, and precision searches for physics beyond the Standard Model (BSM)~\cite{hutzler2020polyatomic, kozyryev2017precision, JilaEDM, ACME2014}.  The depth and breadth of each of these applications are determined by both coherence time and molecule number. In many proposed experiments, trapping molecules in conservative potentials is key to achieving a long coherence time, which enhances statistical sensitivity~\cite{hutzler2020polyatomic, kozyryev2017precision}. Furthermore, a small trap volume enables precise control of environmental fields, which reduces systematic errors. Several molecular species have been cooled and trapped in sub-millimeter optical dipole traps (ODT)~\cite{HallasOpticalTrappingCaOH, anderegg2018laser, sawaoka2026optical, langin2021SrFODT} and tweezer arrays~\cite{VilasCaOHTweezerArray, BaoDipolarSpinExchange, HollandOnDemandEntanglement, liu_molecular_assembly_2019, ruttley_enhanced_quantum_control_2024} for a variety of quantum science applications. 

Due to the complexity of molecular laser cooling and constrained production methods, the typical number of molecules in a MOT ($N_\text{MOT}$) --- the workhorse precooling method used to load ODTs --- remains several orders of magnitude lower than those in state-of-the-art experiments with atoms. This has limited the potential of ultracold molecules: precision measurements with trapped molecules rely on large numbers to decrease uncertainty, while quantum simulation experiments depend on high number densities to probe many-body physics and achieve quantum degeneracy. Because of this, several efforts have been made toward improving the number of ultracold molecules in ODTs ($N_\text{ODT}$). In direct laser-cooling experiments, chemical enhancement~\cite{Jadbabaie2020} and high-compression blue and conveyor belt (CB) MOTs~\cite{HallasBlueMOT, sawaoka2026optical} have each increased $N_\mathrm{ODT}$ by close to an order of magnitude. These works also demonstrated record number densities ($n$) in molecular MOTs and ODTs~\cite{jorapur2024HighDensity, YuConveyorBeltMOTCaF, youssef2026EnhancedMOT}, with $N_\mathrm{ODT}$ being improved by two orders of magnitude since the first molecular ODTs with CaF, SrF and CaOH~\cite{HallasOpticalTrappingCaOH, anderegg2018laser, langin2021SrFODT}. Improvement in $N_\text{ODT}$ was also realized by using the MOT itself as a highly sensitive spectroscopic instrument to find laser repumping transitions that increase the photon budget~\cite{Lunstad2026}. Despite these advances, several key milestones in cold molecular quantum science are still out of reach, requiring higher $N_\text{ODT}$.

The dominant remaining loss mechanism when loading a MOT from a molecular beam is related to the geometry of the beam, where its angular spread results in a small intersection with the MOT region. For example, a typical molecular beam has a divergence of ${\sim}1$~sr and propagates a distance of $\sim$$ 1$~m toward a MOT with typical capture radius $\sim $$1~\mathrm{cm}$. This results in only $\sim$$10^{-4}$ of the molecules having a chance to be captured in the MOT, which is several orders of magnitude worse than the other loss mechanisms present in these experiments. This fraction can be increased by collimating the beam with two-dimensional transverse cooling. Sisyphus cooling in particular promises transverse beam temperatures in the sub-Doppler regime, resulting in minimal transverse velocity spreads, smaller effective beam divergence, and it is often proposed as a viable technique to circumvent this large loss mechanism~\cite{Langin_2023_improved_loading}.
Although sub-Doppler transverse cooling was used in preliminary demonstrations of photon cycling for laser-coolable molecular candidates~\cite{KozyryevSrOH, MitraCaOCH3,li2026AsymmetricTopCooling, AugenbraunYbOH}, and as a way to improve flux in untrapped molecular beam experiments~\cite{Alauze_2021_YbF_transverse, vanhofslot2026BaFTC}, it was not found to enhance $N_\text{MOT}$. While prior studies involving Doppler transverse cooling realized order unity improvements in MOT number~\cite{rich2024thesis}, the same work showed a reduction in $N_\text{MOT}$ when applying sub-Doppler transverse cooling. 

Here we report a sub-Doppler transverse molecular beam cooling method that increases the number of optically trapped SrOH molecules by more than an order of magnitude. We explore how this improvement depends on several experimental parameters, including the overall and relative detunings of the transverse cooling lasers, as well as their intensities. Additionally, we characterize key aspects, such as the timing and spatial position of the cooling light, as well as the importance of ``pointing'' the collimated molecular beam after it has been cooled~\footnote{This is discussed in Appendix~\ref{SM:Beam}}. This method is applicable to all current experiments with molecular MOTs, and thus has strong potential for wide impact across quantum science, including precision particle physics and quantum information processing. 
\begin{figure*}[t]
  \centering
  \includegraphics[width = \textwidth]{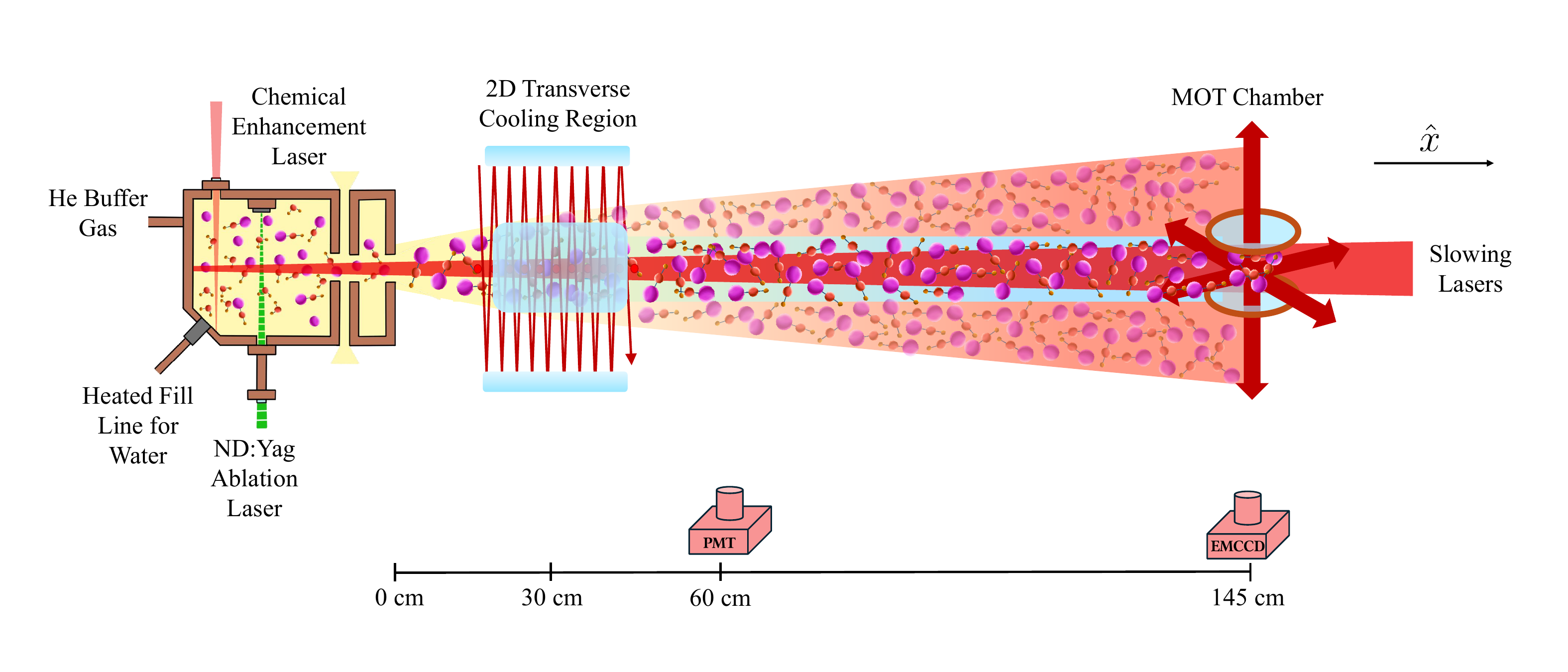}
  \captionsetup{justification=justified, singlelinecheck=false}
  \caption{\justifying Experimental sequence, starting with SrOH production in a cryogenic buffer gas beam (CBGB) cell. The molecules propagate 30~cm and then encounter the 2D transverse laser cooling (TC) region. Timing of laser pulses is set so that after TC they are then laser slowed for subsequent trapping in the MOT. The molecular beam is monitored via fluorescence with a photomultiplier tube (PMT) located 60~cm downstream of the CBGB cell to benchmark molecular production and to measure the molecular arrival time after the Nd:YAG is fired (which produces the SrOH precursor strontium atoms). The MOT and ODT, which are positioned $145~\mathrm{cm}$ downstream of the cell, are imaged on an EMCCD. The faded molecules in the red cone in the figure represent (not to scale) the spread of the beam without transverse cooling applied, where the vast majority of them end up missing the MOT capture region. The opaque molecules within the blue cone represent the increased on-axis molecular flux due to the transverse cooling of the beam.}
  \label{Figure:Schematic}
\end{figure*}

An immediate application of this method will be in precision measurements using SrOH for ultralight dark matter (UDM) and the electron EDM (eEDM) searches. SrOH is currently used to search for UDM by probing for oscillations of the proton-to-electron mass ratio, $\mu=m_p/m_e$. Our results position that experiment to improve its sensitivity to UDM by an order of magnitude over current state-of-the-art limits\cite{CentersBosonicDarkMatter2021}\footnote{This is discussed in Appendix~\ref{SM:UDM}}. Future proposed experiments with SrOH searching for the eEDM can probe for new CP-violating particles with $\gg$ 10 TeV mass. The high number of trapped molecules achieved here is a crucial enabling step. An SrOH-based eEDM search also offers a distinct combination of sensitivities to $d_e$ and $C_S$ in a global analysis with other eEDM experimental results, which could meaningfully constrain or discover sources of new CP-violating physics. Beyond SrOH specifically, a wide range of quantum science applications with other laser cooled molecules will benefit from this method --- the reader is directed to the conclusion of this paper for more discussion. 

\begin{figure}[htbp]
  \centering
  \includegraphics[width=0.95\linewidth]{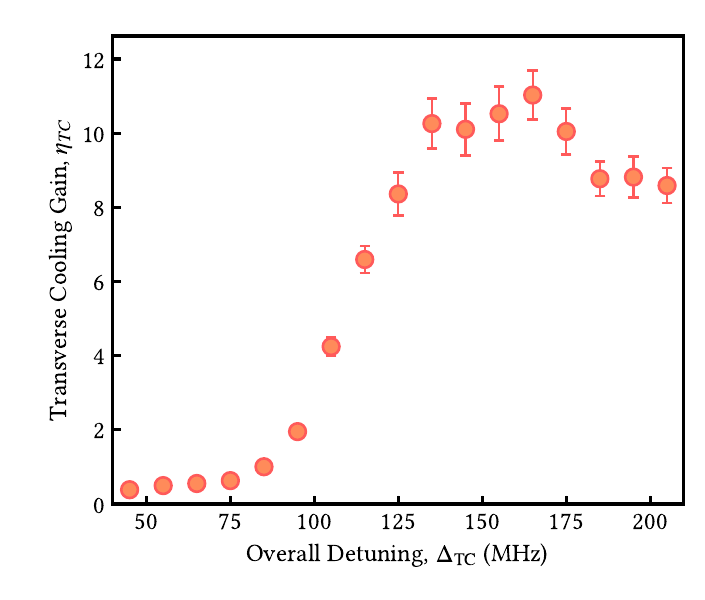}

  \captionsetup{justification=justified, singlelinecheck=false}
  \caption{\justifying Multiplicative increase $\eta_{\text{TC}}$ in $N_\text{MOT}$ as a function of overall detuning ($\Delta_\mathrm{TC}$) of the TC lasers relative to the $J = 3/2$ spin-rotation level. The relative detuning between the (I) and (II) frequency components $\delta_\text{TC}$ in each cooling axis is held fixed to a value of $110~\mathrm{MHz}$. A maximum in $\eta_{\text{TC}}$ occurs near $\Delta_{\text{TC}} \sim \OverallDetuning~\mathrm{MHz}$. Around $\Delta_{\text{TC}} \sim 90~\mathrm{MHz}$, cooling becomes ineffective because the lower-frequency component approaches resonance with the $J = 3/2$ level, disrupting the Sisyphus cooling mechanism.}
  \label{fig:Detuning}
\end{figure}

\begin{figure}[t]
  \centering
  \includegraphics[width=0.46\textwidth]{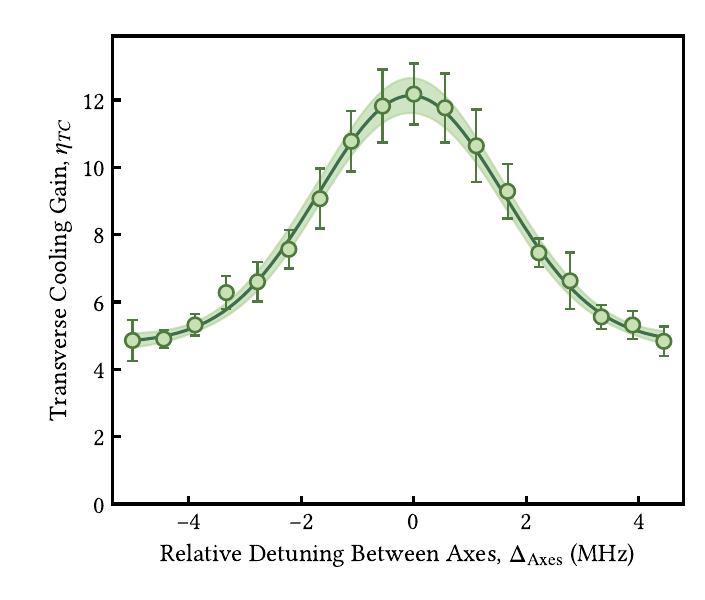}

  \captionsetup{justification=justified, singlelinecheck=false}
  \caption{\justifying
$\eta_{\text{TC}}$ as a function of the relative detuning between the two cooling-axis frequencies, $\Delta_{\text{Axes}}$ (see text).
  }
  \label{fig:Detuning_Axes}
\end{figure}

 \begin{figure*}
  \centering
  \includegraphics[width = 1.05\textwidth]{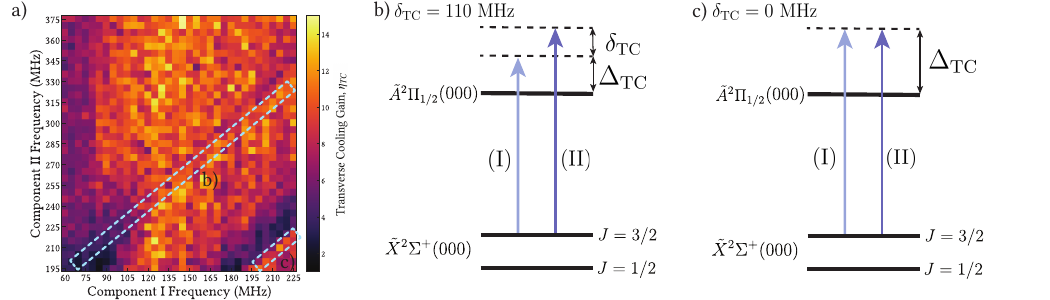}
  \captionsetup{justification=justified, singlelinecheck=false}
  \caption{\justifying a) Increase in MOT number for scans varying $\delta_\text{TC}$, the spacing between frequency components in each axis, holding $\Delta_\text{Axes}$ = 0. The overall structure is reflective of the characteristic blue-detuned sub-Doppler cooling mechanism. b) Within Fig.~$a$, depicted is a slice corresponding to $\delta_\text{TC} = 110$~MHz, as shown in level diagram $b$, which equals the ground state spin-rotation splitting, characteristic of $\Lambda$-enhanced gray molasses cooling. c) Within Fig. $a$, a slice corresponding to $\delta_\text{TC} = 0$~MHz, as shown in level diagram $c$, characteristic of single-frequency cooling.}
  \label{Figure:2DFreq:a}
  \label{Figure:2DFreq:b}
  \label{Figure:2DFreq:c}
\end{figure*}

\section{Experimental Setup}
The laser cooling of SrOH begins with a cryogenically cooled buffer gas beam (CBGB) of SrOH, a well-established technique used in many other direct laser cooling experiments~\cite{HutzlerCBGB, VilasCaOH, Baum2020magneto, BarrySrF, CollopyYO, Padilla-Castillo_AlF}. The beam is radiatively slowed and trapped in a MOT as described in \cite{LasnerMOT}, after which the molecular cloud is sub-Doppler cooled, then compressed in a CB MOT and trapped in an ODT \cite{sawaoka2026optical}. Transverse cooling of the molecular beam is done via Sisyphus cooling in two dimensions as shown in Fig.~\ref{Figure:Schematic}. Two sets of standing waves are orthogonal to the propagation direction of the molecular beam and to each other. These standing waves create spatially varying light shifts such that molecules ride up to the peak of a potential energy hill at an intensity maximum, where they are optically pumped into a dark energy sublevel. Following this, they are spatially remixed and arrive at a potential energy minimum. This cycle results in net kinetic energy removal. Given the type-II laser cooling transition used here with SrOH, where the magnitude of angular momentum is larger in the ground state compared to the excited state, cooling (heating) occurs at an overall blue (red) detuning relative to the main transition. 

In the experiment, the transverse cooling (TC) light is applied 30~cm downstream from the CBGB cell and is turned on simultaneously with ablation for 13~ms thereafter, after which the slowing light is applied. As discussed further in SM~\ref{SM:Beam}, this ``buffer time'' is crucial as the slowing light disrupts the TC mechanism. A separate laser is used for each TC axis. The vertical transverse cooling laser (VTC) laser is supplied by a dedicated 688~nm sum-frequency generated (SFG) laser that outputs 5~W of power. For the horizontal transverse cooling (HTC) axis, the light is produced by the laser that is used for white-light longitudinal slowing of the molecular beam, using an AOM to switch quickly between the applications. The HTC laser is beat-note locked to within 70~kHz of the VTC laser, which is in turn absolutely locked to within 5~MHz of a lithium reference using a wavemeter.

Each cooling axis comprises two frequency components generated by an acousto-optic modulator (AOM) split by $\delta_\mathrm{TC} = 110~\mathrm{MHz} $ to address the ground state spin-rotation (SR) energy levels. These components are labeled (I) and (II) for the lower and higher frequencies, respectively, and are combined on a polarizing beam-splitter. For each axis, frequency components (I)/(II) have linear polarizations orthogonal to each other, as well as to the propagation direction of the particular transverse cooling axis. Furthermore, a static magnetic field of $1.5$~G/cm is applied at a $45^\circ$ angle relative to each coordinate axis, where the molecular beam travels in the $\hat{x}$ direction as shown in Fig.~\ref{Figure:Schematic} \footnote{We have not observed an effect from the static magnetic field in 2D cooling. For 1D cooling, some magnetic field appeared beneficial and we retain that optimal 1D value in our 2D cooling.}. The light is aligned into a photonic crystal fiber and delivered to the TC region with a Gaussian beam waist of $\BeamSize$~mm. The laser beam then propagates through a 3x beam expander, followed by a 10x cylindrical telescope resulting in expansion in the direction orthogonal to the propagation of the molecular beam in order to increase the cooling volume. 

These laser beams with dimensions of $30\times3$ mm  propagate through the TC region. Optical access in both the VTC and HTC directions is provided with four mounted anti-reflection (AR) coated windows optimized for 688 nm light. The light is multi-passed $\sim$$20$ times over a distance of 50~mm between two mirrors to create a standing wave. Both VTC and HTC beams each have a power of $2.5~\mathrm{W}$. The power is split between the spin-rotation frequency components such that $\TCBeamsratio\%$ of the power is allocated to component (II), and the rest to component (I). These powers correspond to intensities of $I_{TC} = \TCIntensity~\text{W/cm}^2$ resulting in $I_{TC}/I_\text{sat} \sim \TCIoverIsat$, where $I_\text{sat} \sim \SrOHIoverIsat $~mW/$\text{cm}^2$ is the saturation intensity for the main cooling transition in SrOH.

In addition to the main cooling light, three lasers are added in the TC region to repump the dominant decays to unaddressed vibrational ground states, specifically the $\X(100), \X(200)$ and $\X(010; N = 1)$ states. This is achieved through the following repumping transitions: $\X(100) \rightarrow \B(000)$ at 631~nm, $\X(200) \rightarrow \B(100)$ at 631~nm, and $\X(010; N = 1) \rightarrow \B(000)$ at 624 nm. The repumping lasers are combined with the TC laser on a long-pass dichroic mirror before the cylindrical lens in each axis, resulting in the ability to scatter $\sim$$900$ photons before appreciable loss to vibrational dark states~\cite{Lunstad2026}. 

\section{MOT and ODT Number Enhancement}

The TC signal was first observed through imaging of the molecular beam shape on an EMCCD camera 60~cm from the CBGB cell, and optimized for spatial beam compression when cooling (and beam expansion when in the heating frequency configuration), as described in Appendix~\ref{SM:Beam}. The position of the compressed molecular beam was shifted to the geometric center of the beamline by altering its pointing after collimating during TC via tilting the TC standing wave relative to the molecule beam propagation direction. This process was repeated once more, but while observing the molecular beam in the MOT region. Once the collimated molecular beam was centered in the MOT region, the trapped molecule number was measured with TC on versus off, where the ratio $\eta_\text{TC}$ between the two conditions was maximized by tuning various parameters, including powers and frequencies of the TC lasers as well as the alignment of the standing waves. Following this optimization, the improvement in the number of molecules trapped in the MOT ($N_\text{MOT}$) and ODT ($N_\text{ODT}$) due to TC was measured. The average multiplicative increase is observed to be $\eta_{\text{TC}} \sim \FactorIncrease$, resulting in  $N_\text{MOT} = \NewMOT(\NewMOTUncert)\times10^5$ and $N_\text{ODT} = \NewODT\left(\NewODTUncert\right)\times10^4$, where the ODT loading efficiency is limited by two-body collisions as discussed below. The value of $\eta_\mathrm{TC}$ can vary between a factor of $\sim  9-15$ due to variation in CBGB molecular beam dynamics. In the ODT, the trapped molecules are imaged after being held for $150~\mathrm{ms}$ to allow for untrapped molecules to escape the imaging region. These realized values of  $N_\text{MOT}$ and $N_\text{ODT}$ represent over an order of magnitude improvement above the previously reported values in Refs.~\cite{sawaoka2026optical,Lunstad2026}. In that earlier work, the efficiency of conversion from molecules produced in the CBGB to $N_\text{ODT}$ was $\sim$$10^{-5}$. This value of $\eta_\mathrm{TC}=\FactorIncrease$ is consistent with our kinematic simulations of the molecular beam with TC applied $d=30$~cm from the cell, with $\eta_\mathrm{TC}\propto 1/\mathrm{d}^2$; see Ref.~\cite{nasir_thesis} for further details. 

The dependence of the transverse cooling on the overall detuning of the light ($\Delta_{\text{TC}}$) is shown in Fig.~\ref{fig:Detuning}, where $\Delta_\text{TC}$ is the detuning of frequency component (I) relative to resonance for the $J = 3/2$ spin-rotation level for the main cooling transition. The laser frequencies for both TC axes are scanned where $\eta_\text{TC}$ is maximized at $\Delta_{\text{TC}}\sim \OverallDetuning~\mathrm{MHz}$. Furthermore, in two dimensions, the performance of the overall cooling has a strong dependence on the frequency difference between the VTC and HTC lasers. As shown in Fig.~\ref{fig:Detuning_Axes}, the cooling performance is maximized when the two axes have a relative detuning $\Delta_\mathrm{Axes}\sim0~\mathrm{MHz}$. This can possibly be attributed to interference effects between the two coherent axes giving a light intensity boost for greater energy removal or improved remixing of coherent dark states in the ground manifold. This result shows that for a given laser source, it is beneficial to split it into at least two TC dimensions with half the intensity in each axis as opposed to a single axis with full intensity. For example, the MOT improvement from a single axis in this work is $\eta_{\text{TC}_{\text{single}}}\sim2.5$. From Fig.~\ref{fig:Detuning_Axes}, when $\Delta_\mathrm{Axes} = 0~\mathrm{MHz}$, the MOT improvement becomes $\eta_\mathrm{TC}\sim4\eta_{\text{TC}_{\text{single}}}$. As such, even in the power-limited regime where $\eta_\mathrm{TC}\propto I$ (where $I$ is the laser intensity), splitting a given available laser along two orthogonal axes would result in a value of $\eta_\mathrm{TC}$ twice as large despite each single-axis intensity being cut in half.  

Fig.~\ref{Figure:2DFreq:a} shows a two-dimensional scan where both $\Delta_\text{TC}$, the overall detuning of the TC lasers, and $\delta_{\text{TC}}$, the relative detuning between the frequency components present in each cooling arm, are changed and $\eta_\text{TC}$ is measured while fixing $\Delta_\text{Axes} = 0~\mathrm{MHz}$. These data were taken in a configuration where the two cooling lasers generated the two frequency components split on a 50/50 beamsplitter to address both cooling axes, instead of the final configuration where each cooling axis has a dedicated laser. As a result, the frequencies were able to be independently scanned over a large range on the order of $\sim$100~MHz. The structure in Fig.~\ref{Figure:2DFreq:a}a) provides some insight into the underlying sub-Doppler cooling mechanism. In the configuration shown in Fig.~\ref{Figure:2DFreq:a}b), the dashed line around the region indicating $\delta_{\text{TC}} \approx 110$~MHz matches the splitting between the ground state spin-rotation levels very closely. Given that the overall detuning of the lasers is blue, this configuration is gray molasses (GM) cooling using a $\Lambda$-system. Fig.~\ref{Figure:2DFreq:a}c) shows a similar region at a condition where $\delta_{\text{TC}} = 0$, analogous to single-frequency (SF) sub-Doppler cooling. Both of these cooling techniques are well understood and have been demonstrated in several laser cooling experiments \cite{TarbuttEngineerSubDopp, TarbuttEngineerSubDopp, VilasCaOH, DingYOGMC}. 

\begin{figure}[t!]
    \centering
    \includegraphics[width=\linewidth]{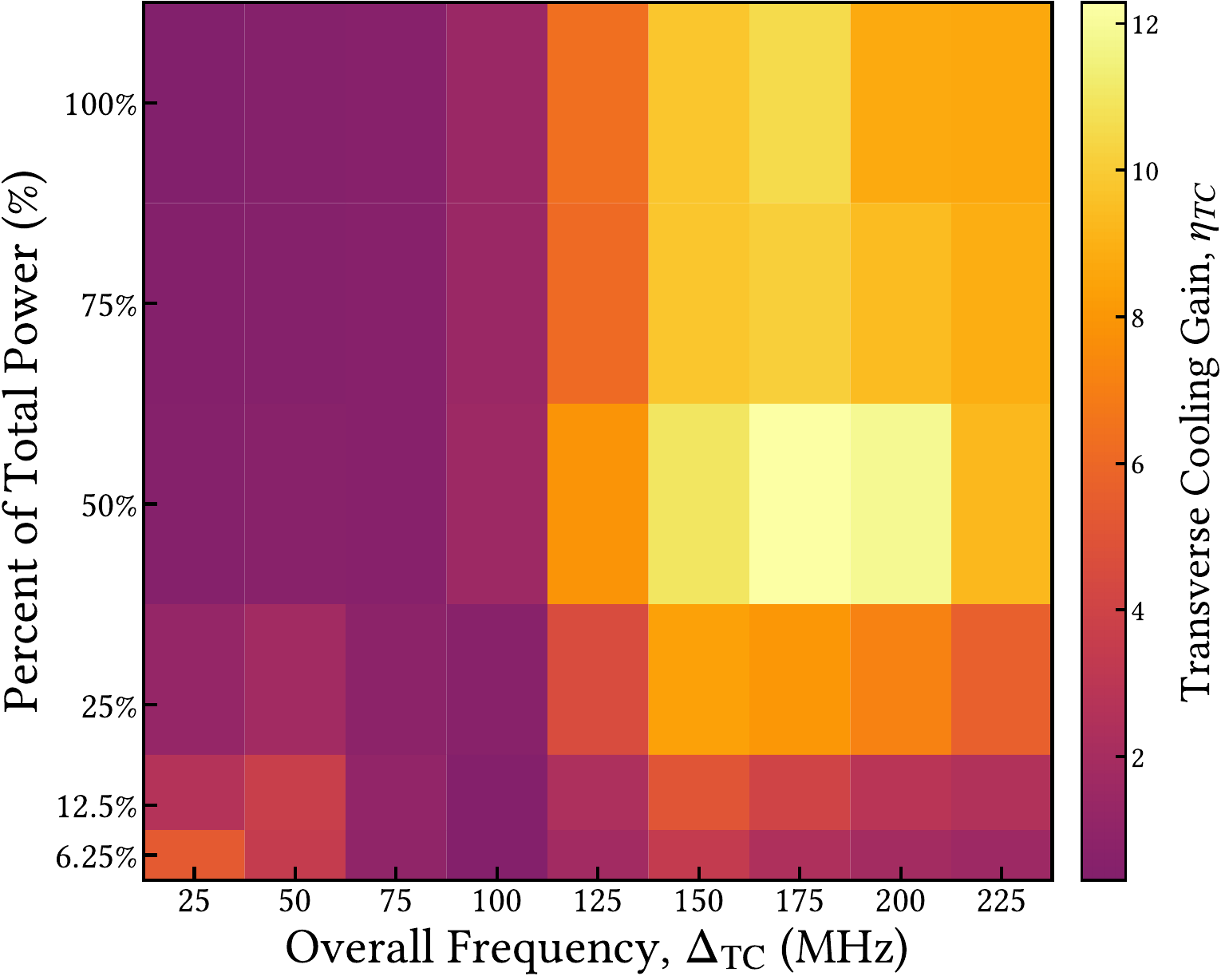}
    \captionsetup{type=figure, justification=justified, singlelinecheck=false}
    \captionof{figure}{\justifying Transverse cooling gain $\eta_\text{TC}$ as a function of the overall frequency $\Delta_\mathrm{TC}$ and the power of the two frequency components with a fixed 3:1 power ratio between components (I) and (II).}
    \label{fig:power}
\end{figure}
For a given axis, we observe that $\eta_\mathrm{TC}$ saturates at a lower power for frequency component (II) compared to (I). As such, we operate with a power ratio $P_\text{(I)/(II)}=3:1$ between the two components. Given that the system forms a $\Lambda$-system, this difference can be explained by seeing (I) as a ``pump'' beam, responsible for setting the standing wave depth which sets the overall cooling capture velocity. The frequency component (II) can be thought of as a "probe" beam, which forms the three-level system and ensures the $\Lambda$-configuration, a process which does not significantly benefit from additional power. This asymmetry in the system comes from (I) being much closer to resonance with the $J = 3/2$ SR level than frequency (II) is to either of the two SR components, consequently resulting in larger light shifts. 

Fig.~\ref{fig:power} shows the dependence of the cooling on the overall power of the TC lasers while holding $P_\text{(I)/(II)}=3:1$. In the current configuration, the power is saturated at $P_\text{TC}\sim \TCPower~\mathrm{W}$ as there is no benefit from further increasing the cooling capture velocity. Kinematic simulations suggest that the cutoff velocity for MOT number gain in this system is at $\sim$1~m/s, since at a distance of 30~cm from the CBGB cell, molecules with larger transverse velocities are cooled at a transverse distance outside of the MOT capture radius. 

\section{Two-Body Collisions in the ODT}

Due to the increased density achieved through TC, we observe molecular two-body loss through SrOH-SrOH collisions in the ODT. Molecules are loaded into a focused, single-beam 1064~nm ODT with a power of $\ODTpower~\mathrm{W}$ and a waist radius of $\ODTwaist~\mu \mathrm{m}$ with a trap depth of $T_\mathrm{depth}= U_0/k_B=\ODTtrapdepth~\mathrm{mK}$. We measure the temperature of the ODT after holding for 300~ms and then releasing and imaging the molecules after variable times for a time-of-flight measurement. $T_\mathrm{ODT} = \ODTtemp(\ODTtempunc)~\mu K$ is measured at various hold times. To calculate the effective volume, we assume that the trap is harmonic such that $V_\mathrm{eff} = (2\sqrt{\pi})^3 \sigma_x \sigma_y \sigma_z$ where $\sigma_i$ are the beam waists in the $x,y,z$ directions and $\sigma_x=\sigma_y = \sigma_r$, the radial width. In our system, $\sigma_r=w \sqrt{T/4 T_\mathrm{depth}} = \radialwidth~\mu\mathrm{m}$ and $\sigma_z=z_R\sqrt{T/2 T_\mathrm{depth}}=\axialwidth~\mathrm{mm}.$ Thus, the effective volume for the \ODTwaist~$\mu$m waist trap is $V_\mathrm{eff} = (2\sqrt{\pi})^3\sigma_r^2\sigma_z = \ODTvol\times10^{-6}~\mathrm{cm}^3$.
\begin{figure*}[t!]
  \centering
  \includegraphics[width=0.8\textwidth]{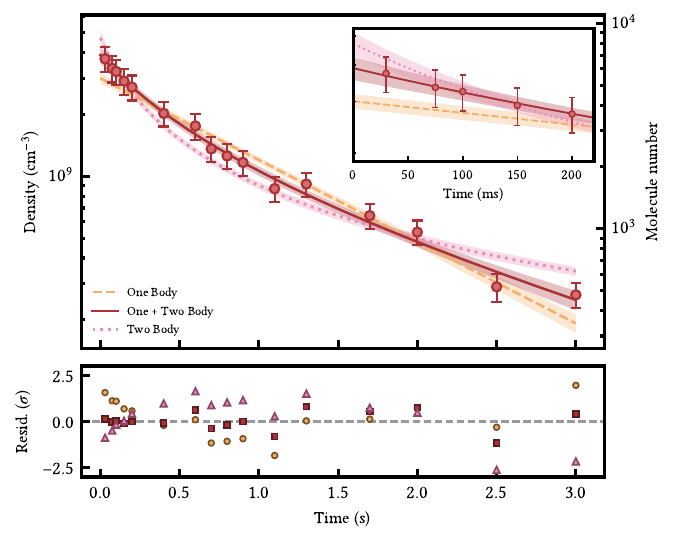}

  \captionsetup{justification=justified, singlelinecheck=false}
  \caption{\justifying
  Measured number of molecules in the ODT as a function of time, fit to a one body, one plus two body, and two body functions, with all parameters free. The one plus two body curve fits best to the data and has a $\tau=\SingleBody$~s, consistent with the one-body lifetime measured in a dilute sample, where collisions are negligible. This fit results in a two-body rate constant of $\beta\sim\Beta\times10^{-10}~\mathrm{cm}^3/\mathrm{s}$ with an estimated systematic uncertainty of a factor of $\sim \!2$, surmised from variations in fit $\beta$ values as described in the text. The normalized residuals for the three fits are shown, where the circles represent the one body fit, the squares the one plus two body fit, and the triangles the two body only fit. The structure of the residuals show that the fit is best for the one plus two body case. Fitting to one plus two body with the one body rate constant set to that measured with dilute samples yields the same $\beta$ within error.
  }
  \label{fig:collisions}
\end{figure*}

We measure the density of molecules in the ODT as a function of hold time as shown in Fig.~\ref{fig:collisions} and fit a two-body loss curve. The fit residuals are smaller for a one plus two body fit compared to separate one or two body fits. The fit determines a single body lifetime of $\tau_\mathrm{single} = \SingleBody(\SingleBodyUnc)~\mathrm{s}$, in good agreement with our exponential decay trap lifetime measured using a very dilute gas of trapped molecules where two-body loss does not play a role. The fitted two body loss coefficient is $\beta\sim \Beta\times10^{-10}~\mathrm{cm}^3\mathrm{/s}$. One plus two body fits at various starting densities are limited by a systematic deviation in $\beta$ of up to a factor of 2 relative to the mean value based on the deviation of the actual molecule distribution from the uniform-density model employed. Using the measured ODT number of $\NewODT(\NewODTUncert)\times10^4$ after holding for time $t = \ODTHold~\mathrm{ms}$, we estimate an ODT number of $\ODTPrediction(\ODTPredictionPlus)\times10^4$ after a hold time of $t = \ODTPredictionHold~\mathrm{ms}$. In the future, in an experiment with increased ODT laser power and a larger ODT beam waist, we predict that concomitant increases in $N_\text{ODT}$ are possible without any other changes to this system. 

\section{Conclusion}

In conclusion, we have demonstrated an enhanced number of optically trapped SrOH molecules using two-dimensional sub-Doppler transverse cooling of a molecular beam, with $\NewODT(\NewODTUncert)\times10^4$ molecules in an ODT, a $\sim$\FactorIncrease$\times$ gain. In doing so, we have developed a technique to  mitigate the largest loss mechanism in molecular laser cooling experiments, namely the solid angle between the CBGB cell and the MOT region. We have elucidated the underlying cooling mechanism and outlined the key considerations for using this technique to improve the number of trapped molecules. There are a number of avenues to further improve upon these gains, such as transverse cooling closer to the molecular source. In the work presented here, the cooling is done 30~cm from the CBGB cell. According to the simulations validated by our experimental results, decreasing this distance to $\sim$10~cm would allow for an additional order of magnitude improvement. Also, the use of a bow-tie cavity configuration for the transverse cooling light could improve robustness and relax laser power constraints needed for high transverse capture velocity and scattering rate. Lastly, increasing the ODT laser volume and power would mitigate the observed collisional loss, directly producing higher trap numbers.

An initial application of this method will be in precision measurements using SrOH for ultralight dark matter (UDM) and electron EDM (eEDM) searches, as elaborated in the introduction section of this paper. Heavier molecules exist that, compared to SrOH, offer even greater sensitivity to physics beyond the Standard Model (BSM).  However, progress in that direction has been limited  because these molecules are more difficult to laser cool and the most sensitive experiments (e.g. nuclear Schiff moment searches with deformed nuclei) use rare isotopes available in very limited quantities with radiological safety constraints (requiring the use of minute quantities of source material) ~\cite{ArrowsmithKronRadioactiveMolecules2024, YuHutzlerSymmetricTop2021, FleigDeMilleRadiumMolecules2021}. Greater efficiency in cooling from the molecular beam into the MOT, and then transferring into the ODT, is crucial for the success of those experiments.  Experiments with YbF~\cite{LimYbF, Alauze_2021_YbF_transverse, Tarbutt2013} and RaF/RaOH~\cite{GarciaRuizRadioactiveMolecules2020, UdrescuRadiumLaserCooling2024, ConnColdRadioactiveMolecules2025} can be aided by the transverse cooling method developed here. Finally, for quantum many-body experiments with laser cooled molecules, a $\times \!10$ improvement in the highest reported ODT densities, from $n_\text{ODT} = 10^{10}$~cm$^{-3}$~\cite{jorapur2024HighDensity, YuConveyorBeltMOTCaF} to $n_\text{ODT} = 10^{11} - 10^{12}$, would cross the density threshold for runaway evaporation ~\cite{davis1995SodiumEvapCooling, masuhara1988HydrogenEvapCooling}. Combined with microwave shielding~\cite{anderegg2021MicrowaveShielding, karman2018MicrowaveShielding, bigagli2024DipolarMoleculeBEC}, this offers a promising path to obtaining quantum degeneracy with directly laser cooled molecules, including polyatomic molecules --- a key milestone not yet achieved.
\begin{acknowledgements}
The authors thank Avikar Periwal for valuable discussions regarding the cooling mechanism as well as reviewing the manuscript. We additionally thank Arian Jadbabaie and Nathaniel Vilas for important discussions regarding collisions. This work was done at the Center for Ultracold Atoms (an NSF Physics Frontier Center) and supported by Q-SEnSE: Quantum Systems through Entangled Science and Engineering (NSF QLCI Award OMA-2016244), the Alfred P. Sloan Foundation (G-2023-21036), the Gordon and Betty Moore
Foundation (7947), AOARD: Asian Office of Aerospace Research and Development (FA2386-24-1-4070), and AFOSR: Air Force Office of Scientific Research (DURIP FA9550-24-1-0060).
\end{acknowledgements}
\bibliographystyle{apsrev4-2}
\bibliography{sroh}
\begin{center}
\textbf{Supplemental Material}
\end{center}

\section*{SM1: ODT number and lifetime}
\label{SM:Calibration}

\textbf{ODT number measurement via red-MOT recapture imaging (RRI)} ~We measure the ODT molecule number by holding the molecules in the ODT (with only the optical trapping light on) for $\tau_\mathrm{hold} = \text{150 ms}$, then turning on the RF red-MOT to recapture and image them for $100~\mathrm{ms}$, which we refer to as {RRI}. To confirm that any molecules that were not trapped in the ODT completely leave the RF MOT recapture region, we measure the molecule number after loading the ODT for a fixed time, after which we turn off the trap light for some time $\tau_\mathrm{hold}$ and image the leftover molecules. Varying $\tau_\mathrm{hold}$ subsequently informs us about how long the untrapped molecules take to exit the RF MOT recapture region. Setting $\tau_\mathrm{hold} = 150~\mathrm{ms}$ ensures that these molecules have fully fallen out of the region such that the background signal is consistent with zero molecules in the RF MOT at this hold time. Previous work calculated the conversion from the camera fluorescence to the trapped molecule number in the red-MOT \cite{Lunstad2026,LasnerMOT}, and the same calibration can be used to report the ODT number using RRI. Moreover, we measure the recaptured molecule lifetime to be $37(1)~\mathrm{ms}$, which is set by the known finite photon budget and scattering rate of the RF MOT beams. A 100~ms imaging window therefore collects $1 - e^{-100/37} = 93\%$ of this budget, resulting in a small correction of $1.07\times$ on the calibrated ODT number to account for loss during imaging due to the finite molecular lifetime. In this work, we report the number of molecules trapped at the time the image is taken as opposed to the extrapolated peak trapped molecule number as was done in previous work.

\textbf{Lifetime measurement via Gray-Molasses Imaging}~~ In order to measure the ODT lifetime, particularly when investigating two-body collisional loss, it is important to image the trapped molecules after short hold times starting at $\tau_\mathrm{hold} < 150~\mathrm{ms}$ to fit the relevant decay curves. RRI is not suitable for a short $\tau_\mathrm{hold}$ as it takes $150~\mathrm{ms}$ for the molecules to escape the MOT region. Instead, we employ gray-molasses imaging (GMI), whereby the light is blue-detuned in the $\Lambda$-enhanced GMI configuration as described in \cite{sawaoka2026optical}. It is sufficient to wait for the untrapped molecules to diffuse out of the imaging region as the sub-Doppler cooling does not provide a restoring force. This is achieved with $\tau_\text{hold} = 30~\mathrm{ms}$ while the region of interest (ROI) is selected to maximally zoom in on the trapped molecules in the ODT. For lifetime scans, GMI is turned on for $2~\mathrm{ms}$ and the trap is imaged for various values of $\tau_\text{hold}$. 

In order to determine the molecule number density, the absolute molecule number must be measured. To do this, the ODT fluorescence is measured by switching between GMI for $2~\mathrm{ms}$ and RRI for $100~\mathrm{ms}$ with $\tau_\text{hold} = 150~\mathrm{ms}$, the latter of which is calibrated to give the molecule number as described above. The ratio between these two imaging methods gives the relevant conversion factor for the molecule number measured with GMI. 

\section*{SM2: Transverse Cooling of the Molecular beam}
\label{SM:Beam}

One-dimensional transverse laser cooling (TC) had previously been demonstrated on SrOH to first show laser cooling of a polyatomic molecule \cite{KozyryevSrOH}. Before using two-dimensional TC to improve the trapped molecule number here, we confirmed cooling in each dimension by imaging the molecular beam. These initial tests were also used to study the interaction of laser slowing and transverse cooling as well as the pointing of the collimated molecular beam. We present the results in brief below.

\subsection{Observations of the molecular beam}

The molecular beam is viewed in two locations: Position A - midway on the beamline ($\sim$1~m from the cell, $\sim$70~cm after the TC region) viewed perpendicularly to the direction of the beam and Position B - at the MOT chamber (1.45~m from the CBGB cell (molecular beam source) and 1.15~m after TC) viewed in the same place as the TC is applied. In each case, a small probe laser beam with both the main transition ($\X(000)-\A(000)$ at 689~nm) and first repumping laser ($\X(100)-\B(000)$ at 631~nm) are used to probe the molecular beam. The $\B(000)-\X(000)$ decay at $\sim$$611$~nm is observed for detection.

Specifically for the beam studies described in this section, TC is achieved by one laser, split by a 50:50 beamsplitter before being expanded in one axis and then multipassed in the TC region. This configuration is different from that described in the main body of this paper. The relevant difference is that the setup described here allows for TC to be applied in \textit{both} axes along with slowing.

For these studies, the limiting aperture for the beam is a UHV shutter with a 14~mm diameter aperture (VS14E1T0L-EC). For the data presented in the main text, the UHV shutter is replaced with a model with a larger, 25~mm diameter (VS25S1T0) where the limiting aperture is now the plate to which the UHV shutter is mounted which has a diameter of 18~mm. These apertures are significantly larger than those used in previous work with SrOH~\cite{KozyryevSrOH} and YbOH \cite{AugenbraunYbOH} to allow the capture and observation of molecules with higher initial transverse velocity. 

\begin{figure}
  \centering
  \includegraphics[width = \columnwidth]{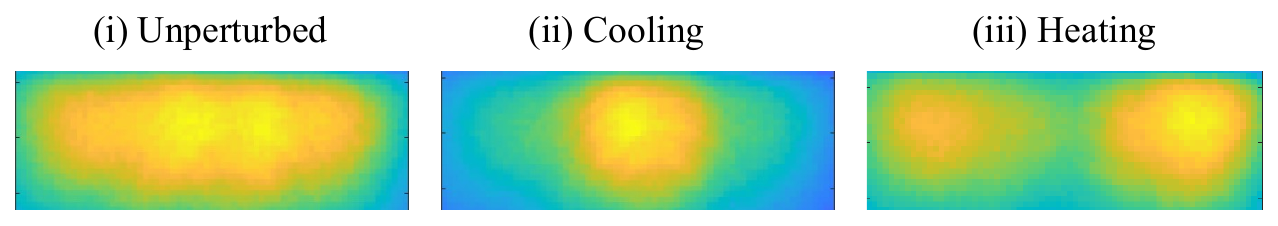}
  \captionsetup{justification=justified, singlelinecheck=false}
  \caption{\justifying Image of the molecular beam observed $\sim$1~m from the cell (70~cm from TC cooling region) on an EMCCD on a normalized color scale, showing both the cooling and heating effects with the laser detuned blue or red, respectively.}
  \label{Fig:BeamProfile}
\end{figure}

Once aligned, the laser frequency for each axis is swept from red to blue detuning relative to the main cooling transition. As a result, both heating and cooling features are observed as shown in Figure~\ref{Fig:BeamProfile}. The molecules are cooled in each axis, where the cooling is confirmed separately in Position A before the camera is moved to Position B, where both axes are cooled concurrently. 

\begin{figure}
  \centering
  \includegraphics[width = .5\columnwidth]{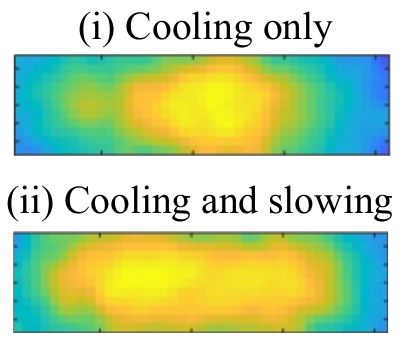}
  \captionsetup{justification=justified, singlelinecheck=false}
  \caption{\justifying Images of the molecular beam with TC applied ($\sim$70~cm after cooling). In condition (i), only cooling was applied. In condition (ii), TC is applied at the same time as the slowing light, leading to a beam with significantly less compression.}
  \label{Fig:SlowingCoolingBeam}
\end{figure}

\begin{figure}
  \centering
  \includegraphics[width = .9\columnwidth]{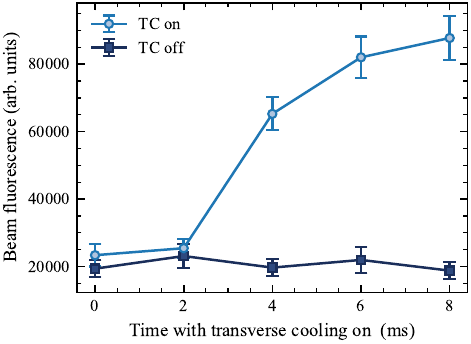}
  \captionsetup{justification=justified, singlelinecheck=false}
  \caption{\justifying Beam fluorescence viewed 70~cm after cooling. The molecules are imaged a total of 11~ms after initial molecule production. TC is applied from 0~ms to a total of 8~ms after the end of enhancement, and the fluorescence increase is observed.}
  \label{Fig:BeamBufferTime}
\end{figure}

\begin{figure}
  \centering
  \includegraphics[width = .9\columnwidth]{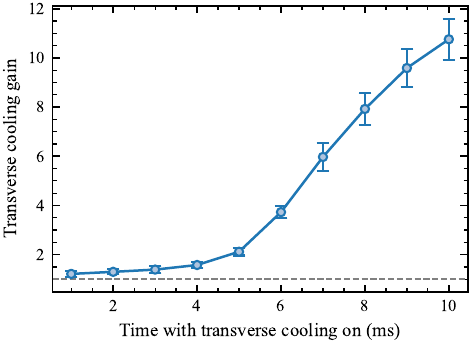}
  \captionsetup{justification=justified, singlelinecheck=false}
  \caption{\justifying With a set total `buffer time' of 10~ms between the end of enhancement and the beginning of slowing, the total time for which TC is on is varied. The highest gain comes with TC on 100\% of this time. Slowing is separately optimized with 10~ms buffer time.}
  \label{Fig:MOTBufferTime}
\end{figure}

When slowing light is applied concurrently with TC, the qualitative indications of Sisyphus transverse heating and cooling are no longer apparent, see Fig.~\ref{Fig:SlowingCoolingBeam}. Measuring the increase in on-axis fluorescence with TC, we scan the start time of the cooling relative to the end of ablation as well as its duration. The beam data show that >6~ms of cooling is required for substantial brightness increase from TC, as seen in Figure~\ref{Fig:BeamBufferTime}. In these tests, TC is turned on 3~ms after the Nd:YAG trigger where the `enhancement laser' is on for the initial 3~ms to drive the intercombination line in strontium and enhance SrOH production. These tests were repeated in the MOT as shown in Fig.~\ref{Fig:MOTBufferTime}.

Since the molecular beam still has a forward velocity of $\gtrsim$100~m/s, the >6~ms required is due to the combined time for molecules to be extracted from the cell and then travel to the TC region.

When later investigating the effect of the slowing light on the TC and resulting MOT improvement, only the vertical TC axis was used as the same laser generates both the horizontal TC axis and slowing light, which cannot be applied at the same time. In this configuration, the effect of the slowing light in decreasing the efficacy of TC is still clear, see Table~\ref{tab:slowingandcooling}.
\begin{table}

    \begin{tabular}{|c | c|}
        \hline
        \textbf{Condition} & \textbf{MOT gain}\\
        \hline \hline
        Transverse cooling off & 1 \\ \hline
        Horizontal transverse cooling on & 2.4(1) \\ \hline
        Horizontal transverse cooling on with slowing & 1.6(1) \\ \hline
    \end{tabular}
    \captionsetup{justification=justified, singlelinecheck=false}
    \caption{\justifying Average molecule number under different conditions. The overall MOT increase is higher when TC is applied alone compared to when TC and the slowing light are both on.}
    \label{tab:slowingandcooling}
\end{table}
Thus, the slowing cuts the gain from TC by $\sim$$40\%$. This result implies that a longer beam line would be more beneficial for MOT loading if transverse cooling will be applied, since the geometric loss will be more than mitigated by TC and the extra distance will allow for good separation of cooling and slowing in time, even if the slowing parameters change and the slowing scattering rate decreases.

\begin{figure}
  \centering
  \includegraphics[width = .9\columnwidth]{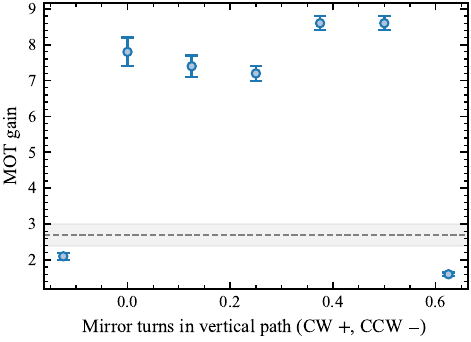}
  \captionsetup{justification=justified, singlelinecheck=false}
  \caption{\justifying MOT gain with both axes in 2D TC, compared to gain from only cooling in the horizontal axis with a factor of $\sim2.7$ improvement as the horizontal dashed line with the variation in the shaded region. Note that the y-axis reflects the maximum gain observed for each position and other adjustments to the multipass did result in lower gains. Additionally, these tests are before final upgrades were made to the system and, thus, show a lower gain.} 
  \label{Fig:AlignmentScan}
\end{figure}

Finally, to see a gain in the number of molecules trapped in the MOT, the collimated molecular beam should be co-linear with the slowing laser beams, as well as intersect with the laser beams that define the MOT region. The pointing of the collimated molecular beam can be controlled by the pointing of the TC laser beams, which in turn controls the angle of the TC standing wave. Since the cooling occurs as molecules climb the potential hills created by the standing wave, the angle of the beams relative to the incoming molecular beam determines the deflection of the molecules. To have zero deflection, the standing wave must be aligned so that it is normal to the molecular beam. In other words, the axis defined by the standing wave is also the one into which the molecules are cooled to $v_\mathrm{transverse}\rightarrow0$~m/s. Thus, the angle of the standing waves (cooling axis) must be controlled to prevent deflection of the molecular beam away from the MOT capture region. While this can be somewhat achieved through careful initial setup, we also scan the angle of the last (or second-to-last) mirror before multipassing the light while recovering the multipass fringes with an upstream mirror. This alignment scan is difficult because maintaining optimal cooling relies on proper alignment of the multipass setup which is affected when the angle of the standing wave is changed. However, with reasonable heuristics for `good' alignment and some care, the scans are repeatable, as shown in Fig.~\ref{Fig:AlignmentScan}\footnote{Despite the reliability of these scans, because other mirrors must be adjusted to maintain the multi-pass alignment, no concrete conclusion is drawn on the absolute angular tolerance of the initial laser beam.}. Additionally in Fig.~\ref{Fig:AlignmentScan}, an incorrect angle on the initial laser beam can deflect the beam from the MOT such that the 2D cooling results in worse MOT gains than the 1D cooling alone, illustrating the negative effects of deflecting the molecular beam away from the MOT.

\begin{figure}
  \centering
  \includegraphics[width = .7\columnwidth]{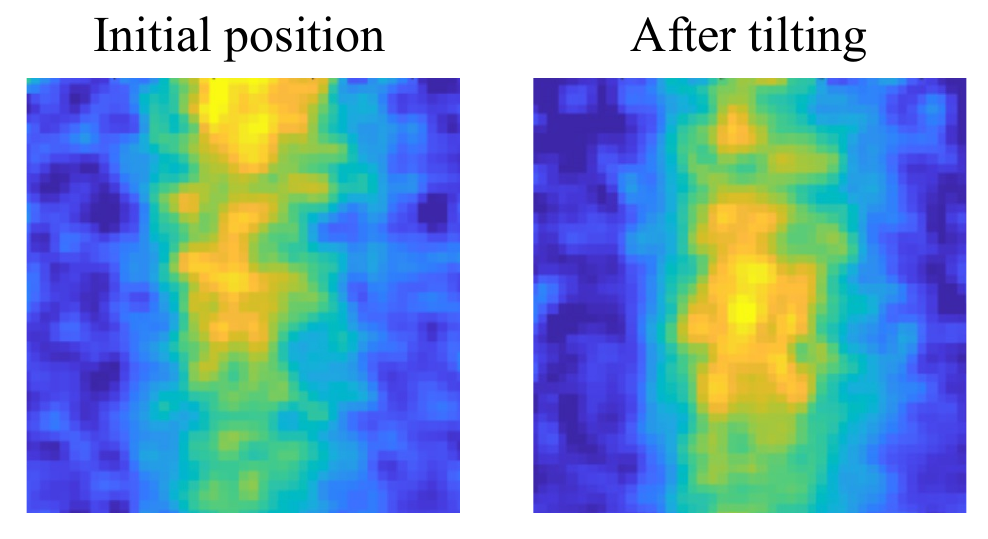}
  \captionsetup{justification=justified, singlelinecheck=false}
  \caption{\justifying Molecular beam fluorescence viewed in the MOT chamber (position B). Cooling is applied in the horizontal axis. Initially, the collimated beam is deflected away from the center of the MOT chamber. After changing the angle of the cooling laser beam, feeding back on the molecular beam position, the cooled molecular beam is centered. Additionally, due to off-diagonal imaging, the SNR is lower than other beam images in this section.}
  \label{Fig:AlignmentPictures}
\end{figure}

The pointing of the molecular beam is also observed in the spatial positioning of the molecular beam itself, in addition to MOT number improvement. Upon initial alignment, the beam is deflected away from the MOT region but can be re-centered through the alignment procedure described above, as seen in Figure~\ref{Fig:AlignmentPictures}.

\section*{SM3: UDM sensitivity estimation}
\label{SM:UDM}

SrOH possesses near-degenerate vibrational levels of differing character, allowing for microwave rovibrational transitions with high sensitivity to variations of the proton-to-electron mass ratio $\mu$~\cite{kozyryev2021enhanced}. Using our measured molecular structure~\cite{Lunstad2026} and the achieved molecule number in our ODT, we project the dark-matter sensitivity via least-squares spectral analysis (LSSA).

Since the previously published sensitivity plots of SrOH to UDM~\cite{kozyryev2021enhanced}, we have measured additional excited vibrational levels in the $\X$ state, allowing better fits of vibrational constants used to estimate the sensitivity as described in the aforementioned Ref.~\cite{kozyryev2021enhanced}. The vibrational energies of SrOH (as a linear triatomic molecule) are given by~\cite{Bernath2005}:
\begin{align}
&G(v_1,v_2,\ell,v_3) 
  = \sum_{i=1}^3 \omega_i\left(v_i+\frac{d_i}{2}\right)+ g_{22}\ell^2\notag\\
    &+ \sum_{i=1}^3\sum_{j\geq i}x_{ij}\left(v_i+\frac{d_i}{2}\right)\left(v_j+\frac{d_j}{2}\right) 
  \quad    \label{eq:vibrationalE}
\\
  &+ \sum_{i=1}^3\sum_{j\geq i}\sum_{k\geq i}x_{ijk}\left(v_i+\frac{d_i}{2}\right)\left(v_j+\frac{d_j}{2}\right)\left(v_k+\frac{d_k}{2}\right).\notag
\end{align}
We fit the equation above for harmonic terms $\omega_1$ and $\omega_2$ and anharmonic terms $x_{11},~x_{22},~x_{12},~x_{111},~x_{122}$ and $x_{222}$, as well as $g_{22}$ to the observed vibrational states in $\X$. Specifically, we fit \{(100), $(01^10)$, (200), $(02^00)$, $(02^20)$, $(03^10)$, $(03^30)$, $(11^10)$, $(12^00)$, $(12^20)$, (300)\}, adding six additional vibrational levels, already published in Refs.~\cite{LasnerMOT, Lunstad2026}, and four new vibrational constants to the fit since the work in Ref.~\cite{kozyryev2021enhanced}. The new fit values used to estimate our sensitivity are given in Table~\ref{tab:molconstants}. The fit values predict the observed energies to below 0.4~$\wn$, or 12 GHz.  For the specific transition under consideration, we follow a proposed transition in Ref.~\cite{kozyryev2021enhanced}, in keeping with the original proposal and in lieu of further Zeeman and Stark spectroscopy data. 

\begin{table}[]
\resizebox{.8\columnwidth}{!}{%
\begin{tabular}{|c|c|c|}
\hline
\textbf{Constant}   & \textbf{Fit $\left(2\pi c \wn\right)$} & \textbf{$~\mu$ dependence$~$} \\ \hline \hline
$\omega_1$ & 541.38             & $\mu^{-1/2}$     \\ \hline
$\omega_2$ & 378.60             & $\mu^{-1/2}$     \\ \hline
$x_{11}$   & -2.54              & $\mu^{-1}$       \\ \hline
$x_{22}$   & -7.03              & $\mu^{-1}$       \\ \hline
$x_{12}$   & -10.34              & $\mu^{-1}$       \\ \hline
$x_{111}$  & 0.019              & $\mu^{-3/2}$     \\ \hline
$x_{122}$  & 0.971              & $\mu^{-3/2}$     \\ \hline
$x_{222}$  & 0.346              & $\mu^{-3/2}$     \\ \hline
$g_{22}$   & 7.51               & $\mu^{-1}$       \\ \hline
\end{tabular}%
}
\captionsetup{justification=justified,singlelinecheck=false}
\caption{\justifying This table shows the results for our fits of Equation~\ref{eq:vibrationalE} to the observed vibrational levels within $\X$ as well as the dependence of these constants on $\mu$.}
\label{tab:molconstants}
\end{table}

We evaluate the amplitude sensitivity at $\mathrm{SNR}=1$ and rescale it to a $95\%$ confidence-level (C.L.) upper limit. Expressed as a constraint on the linear electron-mass coupling $d_{m_{e}}$ (assuming the scalar couples dominantly to the electron mass, $d_{g}\approx 0$ and saturates the local density $\rho_{\mathrm{DM}}\approx
0.4~\mathrm{GeV\,cm^{-3}}$), this projects up to an order-of-magnitude improvement over existing clock-comparison limits~\cite{Sherrill2023, Kobayashi2022, kennedy2020precision} in the low-mass regime for one month of integration. This includes the additional application of a stochastic-degradation factor of $\sim$3 to account for the stochastic fluctuation of the bosonic dark matter~\cite{CentersBosonicDarkMatter2021}. In the low-mass regime, the total measurement time $T$ is far shorter than the field coherence time $\tau_{c}$, and a single experiment samples only one realization of the Rayleigh-distributed dark matter field amplitude, which can be below its rms value $\Phi_{\mathrm{DM}}$. This factor applies to the linear coupling and matches the treatment adopted for the clock-comparison limits, to facilitate a direct comparison. 

\begin{figure}
    \centering
    \includegraphics[width=\columnwidth]{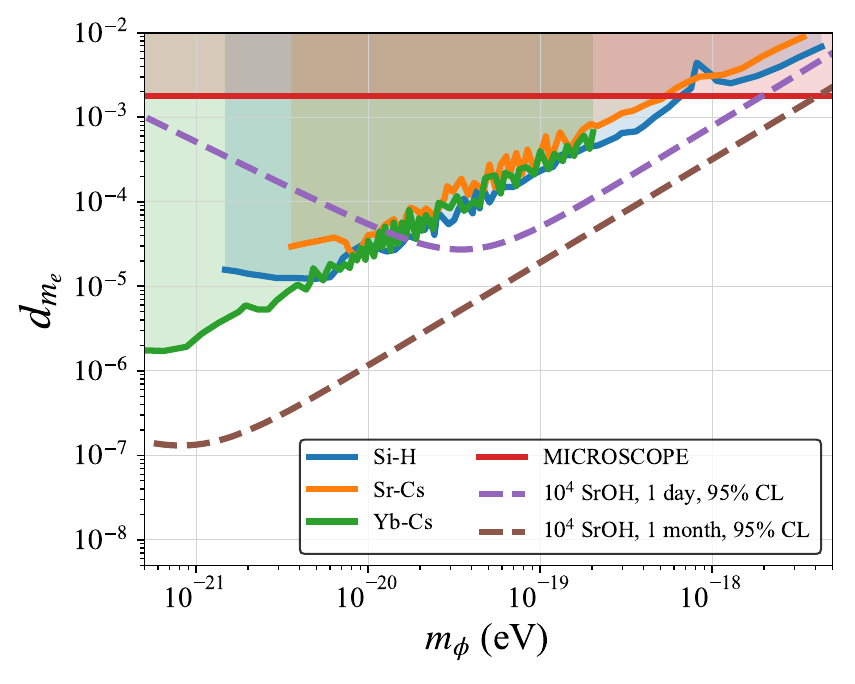}
    \captionsetup{justification=justified, singlelinecheck=false}
    \caption{\justifying Projected $95\%$~C.L. sensitivity of laser cooled SrOH to the linear electron mass coupling $d_{m_{e}}$ of ultralight scalar dark matter versus boson mass $m_{\phi}$. Dashed curves are our projections for $10^{4}$ trapped molecules with one day (purple) and one month (brown) of integration, obtained from an LSSA estimate and including stochastic-degradation factor ~\cite{CentersBosonicDarkMatter2021}. Solid curves are existing clock-comparison limit: Sr/Cs and Yb/Cs~\cite{Sherrill2023, Kobayashi2022} and Si/H cavity-clock comparisons~\cite{kennedy2020precision}, which adopt the same $95\%$~C.L. and stochastic-degradation factor. The red line is the MICROSCOPE equivalence-principle bound~\cite{hees2018violation}, which is independent of the dark-matter amplitude assumption.}
    \label{fig:sroh_dm}
\end{figure}

\end{document}